\documentclass[%
 aip,
 amsmath,amssymb,
 reprint,%
]{revtex4-2}

\usepackage{graphicx}
\usepackage{dcolumn}
\usepackage{bm}
\usepackage{xcolor}

\begin{document}

\preprint{AIP/123-QED}

\title[Lattice-induced spin dynamics in Dirac magnet CoTiO$_3$]{Lattice-induced spin dynamics in Dirac magnet CoTiO$_3$}

\author{A. Baydin}
    \thanks{These authors contributed equally to this work}
    \email{baydin@rice.edu}
    \affiliation{Department of Electrical and Computer Engineering, Rice University, Houston, Texas 77005, USA}%
    \affiliation{Smalley-Curl Institute, Rice University, Houston, Texas, 77005, USA}
    \affiliation{Rice Advanced Materials Institute, Rice University, Houston, Texas, 77005, USA}

\author{J. Luo}%
    \thanks{These authors contributed equally to this work}
    \affiliation{Department of Materials Science and NanoEngineering, Rice University, Houston, Texas 77005, USA}
    \affiliation{Applied Physics Graduate Program, Smalley-Curl Institute, Rice University, Houston, Texas 77005, USA}
\author{Z. He}%
    \affiliation{Department of Physics, University of North Texas, Denton, TX 76203, USA}
\author{J. Doumani}
\affiliation{Department of Electrical and Computer Engineering, Rice University, Houston, Texas 77005, USA}
\affiliation{Applied Physics Graduate Program, Smalley-Curl Institute, Rice University, Houston, Texas 77005, USA}
\author{T. Lin}
    \affiliation{Department of Materials Science and NanoEngineering, Rice University, Houston, Texas 77005, USA}
    \affiliation{Applied Physics Graduate Program, Smalley-Curl Institute, Rice University, Houston, Texas 77005, USA}
\author{F. Tay}
    \affiliation{Department of Electrical and Computer Engineering, Rice University, Houston, Texas 77005, USA}%
    \affiliation{Applied Physics Graduate Program, Smalley-Curl Institute, Rice University, Houston, Texas 77005, USA}
\author{J. He}
\affiliation{Department of Mechanical Engineering, The University of Texas at Austin, Austin, Texas, 78712, USA 
}
\author{J. Zhou}
\affiliation{Department of Mechanical Engineering, The University of Texas at Austin, Austin, Texas, 78712, USA 
}
\affiliation{Center for Dynamics and Control of Materials, The University of Texas at Austin, Austin, Texas 78712, USA 
}
\author{G. Khalsa}
\affiliation{Department of Physics, University of North Texas, Denton, TX 76203, USA}

\author{J. Kono}
\affiliation{Department of Electrical and Computer Engineering, Rice University, Houston, Texas 77005, USA}%
\affiliation{Smalley-Curl Institute, Rice University, Houston, Texas, 77005, USA}
\affiliation{Rice Advanced Materials Institute, Rice University, Houston, Texas, 77005, USA}
\affiliation{Department of Materials Science and NanoEngineering, Rice University, Houston, Texas 77005, USA}
\affiliation{Department of Physics and Astronomy, Rice University, Houston, Texas 77005, USA}

\author{H. Zhu}
\email{hz67@rice.edu}
\affiliation{Department of Electrical and Computer Engineering, Rice University, Houston, Texas 77005, USA}%
\affiliation{Smalley-Curl Institute, Rice University, Houston, Texas, 77005, USA}
\affiliation{Rice Advanced Materials Institute, Rice University, Houston, Texas, 77005, USA}
\affiliation{Department of Materials Science and NanoEngineering, Rice University, Houston, Texas 77005, USA}
\affiliation{Department of Physics and Astronomy, Rice University, Houston, Texas 77005, USA}


\begin{abstract}
Spin-lattice coupling is crucial for understanding the spin transport and dynamics for spintronics and magnonics applications. Recently, cobalt titanate (CoTiO$_3$), an easy-plane antiferromagnet, has been found to host axial phonons with a large magnetic moment, which may originate from spin-lattice coupling. Here we investigate the effect of light-driven lattice dynamics on the magnetic properties of CoTiO$_3$ using time-resolved spectroscopy with a THz pump and a magneto-optic probe. We found resonantly driven Raman active phonons, phonon-polariton-induced excitation of the antiferromagnetic magnons, and a slow increase in the polarization rotation of the probe, all indicating symmetry breaking that is not intrinsic to the magnetic space group. The temperature dependence confirmed that the observed spin dynamics is related to the magnetic order, and we suggest surface effects as a possible mechanism. Our results of THz-induced spin-lattice dynamics signify that extrinsic symmetry breaking may contribute strongly and unexpectedly to light-driven phenomena in bulk complex oxides. 
\end{abstract}

\maketitle
\section{Introduction}

Spin-phonon coupling is crucial for understanding the spin dynamics and transport properties in modern spintronics and solid-state spin-based quantum technologies. Through hybridization of spin and lattice dynamics, novel quasiparticles are formed between spin waves (magnons) and lattice vibrations (phonons) that open the door to new strategies for magnetic control.~\cite{cornelissen_nonlocal_2017,zhang_thermal_2019,streib_magnon-phonon_2019,sivarajah_thz-frequency_2019,bozhko_magnon-phonon_2020,hioki_coherent_2022,luo_evidence_2023}. These hybrid excitations inherit the optical response of phonons and the nonreciprocal properties of magnons, offering pathways for low-loss information transduction and manipulation of energy transport at the nanoscale. Recent advances have demonstrated the potential of engineered artificial lattices, such as hybrid magnon-phonon crystals, to precisely tune these interactions by exploiting symmetry-breaking effects and band structure engineering for phonons and magnons~\cite{LiaoEtAl2024npjS}. The magnon-phonon interaction can also be observed without resonance in nonlinear frequency mixing~\cite{nova_effective_2017}. Cobalt titanate (CoTiO$_3$), a quantum antiferromagnet with a honeycomb lattice structure, has recently been predicted as a model system to explore magnon-phonon coupling. CoTiO$_3$ hosts topological magnons with Dirac-like dispersion~\cite{Yuan_2020,elliot2020visualization} and exhibits phonons with large magnetic moments due to strong spin-orbit coupling and trigonal lattice distortion~\cite{chaudhary2023giant,LujanEtAl2024PNAS,MaiEtAl2024}. Experimental studies have revealed that spin-orbit excitons in CoTiO$_3$ hybridize with phonon modes, imparting a magnetic moment and enabling magneto-phononic effects observable through Raman spectroscopy~\cite{LujanEtAl2024PNAS}. However, the microscopic mechanisms of this coupling have not yet been verified by time-resolved dynamics. 

Here, we have investigated the effect of phonon-polariton excitation on the magnetization of CoTiO$_3$ using time-resolved polar magneto-optic spectroscopy under multicycle terahertz (THz) pumping. Our results showed a fast, resonant phonon-induced birefringence, a phonon-polariton driven excitation of the antiferromagnetic magnon mode, and a slow, non-resonant but polarization-dependent magneto-optic rotation. Such spin dynamics are somewhat unexpected since the material should have no net out-of-plane magnetic moment, suggesting that the lattice excitation should not generate a polar (out-of-plane) magneto-optic response. The THz-induced polarization rotation is highly temperature dependent and correlates with the magnetic order, and thus must originate from spin-lattice coupling. Through symmetry analysis, we postulate that the sample's non-uniform excitation due to the finite penetration depth of THz radiation and/or a surface-induced net magnetization are consistent with our observations. The fact that a homogeneous spin-lattice coupling under the bulk material's symmetry cannot account for such a large spin response illustrates the importance of surface effects and the complexity in analyzing the results in future experiments of phonon-polariton-coupled ultrafast magnetism. 

\section{Results}

\begin{figure}
    \centering
    \includegraphics[width=\linewidth]{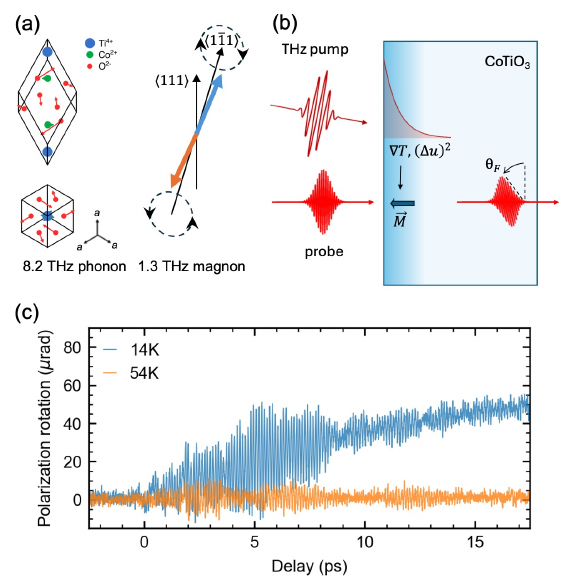}
    \caption{
    (a) [Left]Schematic representation of ion displacements in the Brillouin zone center for $E_g$ phonon at 8.2\,THz in the two projections according to the DFT calculations as adapted from Ref.~\cite{DubrovinEtAl2021JoAaC}. The arrows illustrate the relative amplitudes of vibrations. 
    [Right] Schematic of 1.3\,THz magnon precession around the $a$ axis. Magnetic moments with opposite directions in adjacent honeycomb layers (ab plane) are indicated by orange and blue arrows. 
    \textcolor{black}{(b) Schematic diagram of the experiment. A slab of CoTiO$_3$ is excited by the THz pump, which creates a thermal gradient resulting in magnetization in the region of the pump absorption profile. The probe polarization is then affected by the induced magnetization. }
    (c) Polarization rotation of the probe pulse as a function of time for two different temperatures. The pump frequency and fluence were around 8\,THz and 0.18\,mJ/cm$^2$, respectively. }
    \label{fig:excitations}
\end{figure}

CoTiO$_3$ has an ilmenite crystal structure (space group R$\overline{3}$). The structure can be visualized as alternating layers of Co and Ti ions arranged in a honeycomb lattice slightly buckled along the $c$-axis in an ABC stacking sequence. Each Co$^{2+}$ ion is surrounded by a trigonally distorted octahedral cage of O$^{2-}$ ions. Similarly, each Ti$^{4+}$ ion is surrounded by six O$^{2-}$ ions, forming a TiO$_6$ octahedron.
These CoO$_6$ and TiO$_6$ octahedra are face-sharing. The lattice supports 8 IR active phonons~\cite{DubrovinEtAl2021JoAaC}. A schematic of the ionic displacement for one such phonon, which is relevant in this study, is shown in Fig.~\ref{fig:excitations}(a, Left). It is a doubly degenerate $E_g$ mode with oxygen atoms moving oppositely to the titanium and cobalt atoms, suggesting large dipolar activity. 

Below the Néel temperature of $T_\mathrm{N} = 38$\,K, CoTiO$_3$ exhibits an antiferromagnetic ordering. The Co$^{2+}$ spins align ferromagnetically along the $\langle1\bar{1}0\rangle$ direction within the honeycomb layers and antiferromagnetically between the layers~\cite{ChoeEtAl2024PRB}. The magnetic order greatly reduces the symmetry to a magnetic space group $P\bar{1}.1'c$ (2.7), potentially enabling rich spin-lattice coupling terms. The magnetic order results in 4 magnon branches. The spin precession for the magnon with frequency at 1.3\,THz at low temperature is drawn in Fig.~\ref{fig:excitations}(a, right) with a net magnetic dipole oscillating along the $\langle11\bar{2}\rangle$ direction. 

We performed THz pump-optical probe spectroscopy on a bulk crystal of CoTiO$_3$ polished to \~40 $\mu$m thickness. The setup is schematically shown in Fig.~\ref{fig:excitations}(b). We generated strong, frequency-tunable, and narrowband multicycle THz pulses by chirped-pulse difference frequency generation in the 4-N,N-dimethylamino-4'-N'-methyl-stilbazolium tosylate (DAST) nonlinear organic crystal. The center wavelengths of the two near-infrared pump pulses before mixing were 1433\,nm and 1500\,nm, providing narrowband pulses centered in the range of 8\,THz to 14\,THz and tunable by their relative delay with a bandwidth-limited duration about 0.4\,ps~\cite{LiuEtAl2017OL,LinEtAl2024ACSP}. To
generate linearly and circularly polarized THz pulses, we split the THz beam and adjusted the phase between the horizontally and vertically polarized components~\cite{LuoEtAl2023S}. The THz pump beam was focused onto the sample with a Gaussian beam diameter of 140\,$\mu$m. The pump pulses were characterized by electro-optic sampling with a GaSe crystal. We probed the material's response with a normal incident probe beam along the $\langle111\rangle$ direction. The output of the Ti:Sapphire amplifier generated a pulse with 800\,nm center wavelength and a pulse duration around 50\,fs at the sample location. The probe pulse was focused to a spot with a diameter of 20\,$\mu$m at the center of the pump pulse. After traversing the sample, the polarization rotation of the probe pulse was analyzed by a standard setup consisting of a phase modulator, a Wollaston prism, and a balanced detector~\cite{LuoEtAl2023S}.

\begin{figure}
    \centering
    \includegraphics[width=\linewidth]{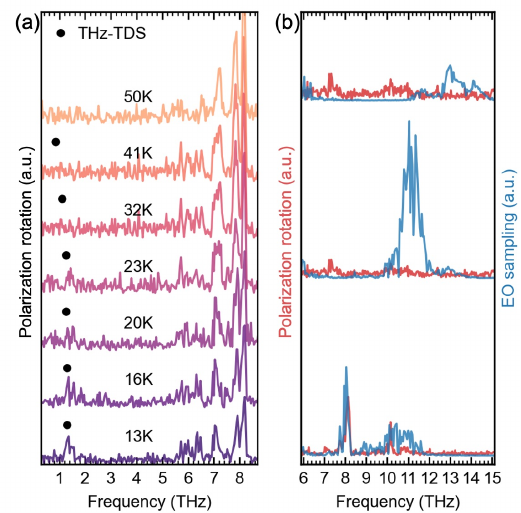}
    \caption{
    (a) Polarization rotation signal in the frequency domain measured as a function of temperature. The black dots indicate the position of the magnon mode as obtained from THz-TDS measurements. 
    (b) Polarization rotation of the transmitted probe pulse in the higher frequency range of the THz pump for three different frequencies, compared with the electric field of the pulse obtained by electro-optic sampling before it impinges on the sample.    
    }
    \label{fig:raw-time-domain}
\end{figure}

Figure~\ref{fig:excitations}(c) shows the polarization rotation as a function of pump-probe delay for two different temperatures, above and below the N\'eel temperature. The pump fluence was 0.18\,mJ/cm$^2$ centered around 8\,THz. The signal at low temperature, in the AFM phase, has two different features: the fast oscillatory response accompanied by a step-like response with a slow rise time. In comparison, the signal above the Néel temperature has no long-term rise; the oscillation pattern is also different. To decompose these features and understand their origin, we first removed the slow response from the trace by using the following fitting equation: 
\begin{equation}
    \theta(t)=A(1-e^{-t/\tau_\mathrm{rise}}) e^{-t/\tau_\mathrm{decay}}, 
\end{equation}
where $\tau_\mathrm{rise}$ and $\tau_\mathrm{decay}$ correspond to rise and decay times of the signal. Then we took a fast Fourier transform (FFT) of the remaining signal at different temperatures across the paramagnetic and AFM phases of CoTiO$_3$, and under different pump frequencies, as shown in Fig.~\ref{fig:raw-time-domain}. For a given pump frequency resonant with the $E_u$ phonon at 8.6 THz, the polarization response in the lower frequency range [Figure~\ref{fig:raw-time-domain}(a)] develops a small peak around 1.3\,THz as the temperature decreases. The frequency of this feature agrees with that of the $\Gamma$-point magnon indicated by the black dots obtained from THz time-domain spectroscopy (THz-TDS)~\cite{ChoeEtAl2024PRB}. 

The high-frequency oscillations peak near 8\,THz remains unchanged with temperature, so it is more likely to have a lattice origin. Figure~\ref{fig:raw-time-domain}(b) compares the high-frequency part of the signal spectra with the pump pulse waveform measured by EO sampling. The gap at around 9\,THz in the EO sampling is due to the inherently low generation efficiency of the DAST crystal and the low detection frequency of the GaSe crystal, both coming from the phonon bands. Meanwhile, the signal was found to concentrate at around 7.2 THz, 8.1 THz, and 10.2\,THz. These apparent resonances do not occur at the actual TO phonon frequencies (8.6\,THz and 12.5\,THz)~\cite{DubrovinEtAl2021JoAaC}, where the incident THz field cannot enter the sample. Instead, they match fairly well with the 3 strong Raman-active phonon modes in this frequency range~\cite{DubrovinEtAl2021JoAaC,ChoeEtAl2024PRB}. 

\begin{figure}[!h]
    \centering
    \includegraphics[width=0.9\linewidth]{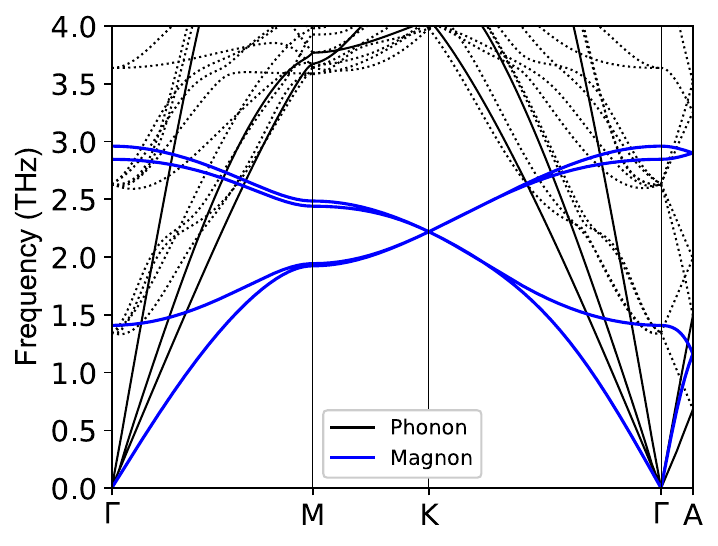}
    \caption{Calculated phonon and magnon dispersion along high symmetry lines in the hexagonal magnetic Brillouin zone. There are 4 magnon modes indicated by solid blue lines (4 magnetic ions in the primitive rhombohedral AFM cell). In the primitive rhombohedral AFM cell, all the zone-center optical phonons are above 5 THz, but zone folding (rhombohedral to hexagonal cell) brings them down, as shown by the dotted lines. Hybridization of magnon and acoustic phonon around 1.5\,THz at finite crystal momentum is possible but is not measured in our experiment.}
    \label{fig:ph_mag_dft}
\end{figure}

The fact that we can observe magnonic oscillations through Faraday rotation and phonon-induced birefringence by THz excitation is puzzling at first glance from a symmetry perspective. The net spin is expected to be in-plane, and the Raman-active modes should be infrared-inactive for a centrosymmetric crystal. The pulse duration is too long for optical impulsive excitation of the high-frequency phonons, which must be driven resonantly by the electric field. Our simulation of the mechanical and magnetic spectra of CoTiO$_3$ confirmed that there are no optical phonons below 5\,THz and no magnons above 3\,THz (see Fig.~\ref{fig:ph_mag_dft} and Table~\ref{tab:dft_phonon}). Therefore, based on the agreement of frequencies, we suggest a phononic origin for the spectral features at 7.2\,THz, 8.1\,THz, and 10.2\,THz, and 
a magnonic origin for the spectral feature at 1.3\,THz. 

\begin{figure}
    \centering
    \includegraphics[width=\linewidth]{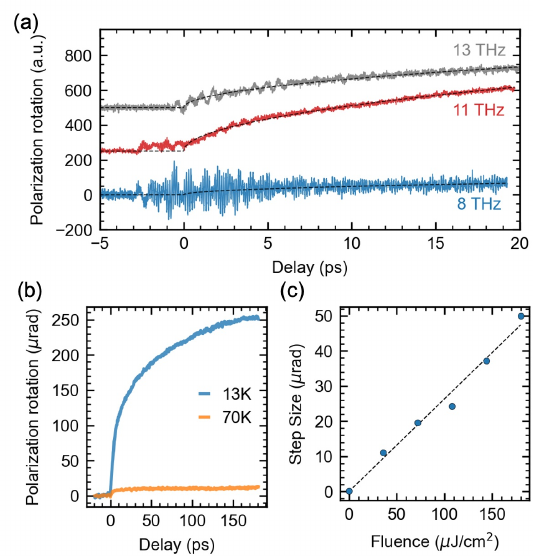}
    \caption{Polarization rotation for different center frequencies of the THz pump maximized when the reflectivity of the crystal is the lowest. 
    (b) The slow rising signal is greatly enhanced below the Néel temperature and not saturated after a long delay. The experimental condition is the same as in Fig.~\ref{fig:excitations}(b) except for being taken with a larger step size. 
    (c) Pump fluence dependence of the rising feature seen in Fig.~\ref{fig:3}. The dashed line is a linear fit to the data. }
    \label{fig:3}
\end{figure}

Note in Fig.~\ref{fig:raw-time-domain}(a) and Fig.~\ref{fig:3}(a) that the low-frequency coherent magnon oscillation is present for all pump frequencies explored in this work. In this frequency region, light-matter interaction with IR-active phonons is expected to be strong: the pulses centered around 8\,THz and 13\,THz excite the TO phonons, and the pulse centered around 11\,THz is near the LO frequency (10.4\,THz) of the $E_u$ mode phonon~\cite{DubrovinEtAl2021JoAaC}. Our data shows that the coherent magnon amplitude is strongest when the TO frequencies are pumped \textcolor{black}{[see Fig.~\ref{fig:3}(a)]}, \textcolor{black}{due to the fact that the dielectric function diverges and the atomic displacement is the largest for a given external field}, consistent with previous work~\cite{nova_effective_2017}. The coupling may arise from a modification of exchange interaction with lattice deformation \textcolor{black}{or phonon displacements}, as demonstrated in CoTiO$_3$ by previous \textcolor{black}{calculations and} experiments~\cite{DubrovinEtAl2021JoAaC, Hoffmann_2021, ChoeEtAl2024PRB}. Given that the 1.3\,THz magnon in fully compensated bulk crystal only exhibits in-plane magnetization, the observation of polarization rotation of a normal-incident probe pulse requires additional symmetry breaking, which can be provided by a surface effect and/or the attenuation of the pump pulses inside the sample, as discussed later.

Next, we focus on the slow dynamics. Figure~\ref{fig:3}(a) compares the time-domain polarization rotation for three excitation pump frequencies, 8\,THz, 11\,THz, and 13\,THz, normalized by total pump power. The slow-rising feature is larger in amplitude at 11\,THz than at the other two frequencies. The excitation is not resonant with any phonons, but the larger response concurs with the range of lower reflectivity, i.e., higher energy deposition into the sample~\cite{DubrovinEtAl2021JoAaC}. Figure~\ref{fig:3}(b) shows a longer time scan above and below the N\'eel temperature. A large rising feature is clearly seen in the magnetic phase, whose lifetime is surprisingly long and beyond our measurement limit of 200\,ps. Such a timescale exceeds both the duration of the excitation pulses and the coherent lifetime of both the phonons and magnons, posing the question of the origin of such a large ferromagnetic-like component in the AFM order. For clarity, we used a larger step for data averaging to smear out the oscillatory part seen in Fig.~\ref{fig:3}(a). We also found that the signal grows linearly with pump power, as demonstrated in Fig.~\ref{fig:3}(c). \textcolor{black}{Finally, we note that the sign of this slow signal does not change when we move to different sample locations for the given crystal.}

\begin{figure}
    \centering
    \includegraphics[width=\linewidth]{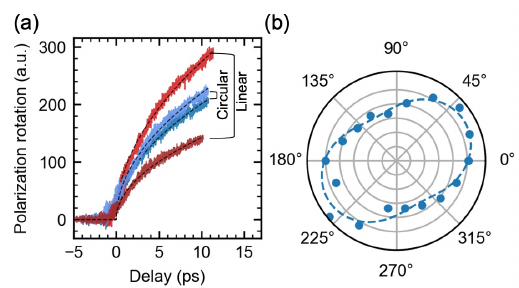}
    \caption{
    (a) The polarization rotation signal for different polarizations of the THz pump.
    (b) The amplitude of the slow-rising feature as a function of the polarization angle of the linear pump polarization.
    }
    \label{fig:anisotropy}
\end{figure}

Figure~\ref{fig:anisotropy}(a) shows the polarization rotation dependence on the type of THz pump polarization (i.e., circular or linear), indicating only a slight change in amplitude of the slow-rising feature between different polarizations. This amplitude is plotted as a function of the polarization angle of the linear pump polarization in Fig.~\ref{fig:anisotropy}(b). Thus, a slight in-plane linear anisotropy is present for THz pump excitation. 

We now attempt to understand the slow dynamics of the polarization rotation signal, which (i) shows up only in the AFM phase (see Fig.~\ref{fig:raw-time-domain} and \ref{fig:3}b); (ii) scales linearly with THz pump fluence (see Fig.~\ref{fig:3}c); (iii) appears in the same direction regardless of linearly or circularly polarized pump (see Fig.~\ref{fig:anisotropy}a), but has (iv) some minor dependence on the angle of the linear polarization of the pump (see Fig.~\ref{fig:anisotropy}b). The first feature, polarization rotation only below $T_N$, reveals its magnetic nature. Therefore, one possible source of the signal is the Faraday effect due to a ferromagnetic component along the propagation direction. The second feature further suggests that this ferromagnetic component originates from a \emph{linear} response.

The E$_g$ modes of CoTiO$_3$ are known to exhibit a large phonon magnetic moment~\cite{ChoeEtAl2024PRB}, and thus, when excited by circularly polarized THz light, regardless of the excitation mechanism, are expected to generate an effective magnetic field and spin polarization~\cite{juraschek_giant_2022,LuoEtAl2023S}. Although this mechanism may account for the small difference in magneto-optic rotation observed with an oppositely circularly-polarized pump in Fig.~\ref{fig:anisotropy}b, it fails to explain the case of the larger response from linearly polarized pump pulses. Another possible origin of the magnetization is the spin Seebeck effect at the crystal surface. At the surface, the symmetry is reduced to $P1$ from $P\bar{1}.1'c$ in the bulk. Therefore, ferromagnetism in any general direction is allowed, but may have the largest component in the plane due to the strong easy-plane exchange anisotropy~\cite{Yuan_2020}. \textcolor{black}{A weak out-of-plane moment is still expected to appear due to the out-of-plane symmetry breaking, although the mechanism that determines the direction of the moment is unclear based on the current data.} The imbalance of opposite spins at the surface will then lead to a net spin current induced by a thermal gradient, i.e. $\vec{J}_s=\kappa\nabla T$, and carried by the magnon~\cite{Adachi_2013}. As the heat from the THz pump diffuses slowly from the surface to the bulk, with a timescale much longer than the pulse duration, the net magnetization will always have the same direction and gradually accumulate. 

With this mechanism, we can approximate the spin accumulation process using a simple 1D model. At time $t=0$, the light pulse instantaneously deposits energy (on the scale of phonon lifetime), and the temperature of the lattice subsystem changes by $\Delta T_{ph}e^{-x/x_0}$, where $x_0$ is the characteristic penetration depth. As the heat capacity of the lattice is likely much larger than the spin subsystem, the lattice acts like a heat bath. Suppose the energy from the lattice continues to flow into the spin subsystem via spin-lattice coupling, which gives a similar spatial profile $T_i + \Delta T_{0}e^{-x/x_0}$, maintains a constant temperature shift $\Delta T_0$, and also a thermal gradient near the surface, by a very crude approximation. \textcolor{black}{$T_i$ is the equilibrium temperature of the system. We have estimated that the temperature increase of the lattice is about 1\,K at the highest fluence, confirming that the system always remains below N\'eel temperature.} 

The 1D heat diffusion equation $d\Delta T/dt=\beta d\Delta T^2/dx^2$ on a semi-infinite domain $x\in[0,+\infty)$ has the solution
\[\Delta T(x,t)=\Delta T_0+\Delta T_0\int_0^\infty G(x,\xi,t)\left(e^{-\xi/x_0}-1\right)d\xi\]
where $G(x,\xi,t)=1/\sqrt{4\pi\beta t}\times\left[\exp(-(x-\xi)^2/4\beta t)-\exp(-(x+\xi)^2/4\beta t)\right]$ is the heat kernel for the Dirichlet boundary condition \cite{Carslaw_1959}. 
Taking the limit $x_0\rightarrow 0$, the solution reduces to a simpler form:
\[\Delta T(x,t)=\Delta T_0\text{erfc}\left(\frac{x}{2\sqrt{\beta t}}\right)\]
where erfc is the complementary error function, but should still preserve the temporal scaling behavior. The thermal gradient near the surface is then
\[\frac{d\Delta T}{dx}\bigg\vert_{x=0}=-\frac{\Delta T_0}{\sqrt{\pi\beta t}}\]
The accumulated magnetization experienced by the probe is proportional to the total flux of spin current generated by the thermal gradient at the surface ($J_s=\kappa d\Delta T/dx|_{x=0}$), which becomes
\[|M|\propto \int_0^t |J_s| dt = \int_0^t \frac{\kappa \Delta T_0}{\sqrt{\pi\beta t}}dt = \frac{2\kappa \Delta T_0}{\sqrt{\pi\beta }}\sqrt{t}.\]
The $\sqrt{t}$ temporal evolution captures the observed long rising time. \textcolor{black}{The magnetization will eventually deviate from the above equation and turn to decay due to spin diffusion at a longer time scale than our experimental range. }To the first order in approximation, the initial temperature rise $\Delta T_0$ at the surface is proportional to the absorbed energy, suggesting a linear dependence of magnetization, and in turn a Faraday rotation, on fluence, which matches well with our observation in Fig. \ref{fig:3}c. In Fig. \ref{fig:3}a, the slow rising curves are fitted with $a\sqrt{t}$ with the relative amplitude $a$ = 36.4, 15.7, 2.7 for the pump frequency of 11 THz, 13 THz, and 8 THz, respectively. The results are consistent with the fact that the total absorbed light power is the largest under 11-THz pump due to smaller reflectance ($R_{11\,\mathrm{THz}}\approx0.4$) than that under 8-THz ($R_{8\,\mathrm{THz}}\approx0.7$) or 13-THz ($R_{13\,\mathrm{THz}}\approx0.8$) pump. The response at 8-THz pump is much smaller, possibly because a large fraction of power is distributed to the 10\,THz band (Fig.\ref{fig:raw-time-domain}b). Finally, a slight anisotropy of the in-plane absorption coefficient caused by the AFM order could result in a small modulation of net spin by the change of pump polarization angle $\phi$ as seen in Fig.~\ref{fig:anisotropy}(b). 

Though the signature of a surface spin Seebeck effect is consistent with all the observed features (i) through (iv), we do not exclude other possible effects. For instance, a strain gradient that transforms like a polar vector, equivalent in symmetry to the thermal gradient, might contribute to the generation of spin current. A pump-polarization-dependent shear strain can contribute a birefringence effect at the probe frequency that might also play a role in the angular variation of polarization rotation in Fig.~\ref{fig:anisotropy}(b). 

\textcolor{black}{The picture discussed above also suggests that in order to distinguish the chiral phonon mechanism from the spin Seebeck mechanism, one may employ thin film samples to modify the thermal gradient and surface properties, although thin film samples also present challenges due to epitaxial stress~\cite{SchoofsEtAl2013JoMaMM}.}

\section{Conclusion}
In summary, we have demonstrated that phonon excitation via phonon-polaritons in CoTiO$_3$ leads to antiferromagnetic magnon dynamics and a slow magneto-optic polarization rotation that evolves over hundreds of picoseconds. These findings provide evidence of light-induced spin-lattice coupling in CoTiO$_3$. The observed temperature dependence of the magneto-optic response confirms the magnetic origin of the observed phenomena. At the same time, symmetry analysis suggests that spatially non-uniform excitation or surface magnetization effects are necessary to explain the slow dynamics of the polarization rotation signal. \textcolor{black}{One interesting extension of the current study is the investigation of thin film samples, where the effect of light-induced temperature gradient could be removed, which in turn can help unveil the alternative physics of chiral phonons.} Our results underscore the importance of carefully interpreting light-driven spin phenomena in complex oxides. This work opens avenues for engineering ultrafast magnetic control via lattice degrees of freedom, with implications for spintronic and magnonic device platforms based on correlated oxides.

\section{Appendix}
\subsection{First-principle calculations}
We carry out first-principle studies in the VASP code \cite{vasp_1,vasp_2,vasp_3,vasp_4}, with projector augmented wave formalism \cite{PAW} and PBEsol exchange-correlation functional \cite{PBEsol}. The kinetic energy cutoff is set at 520 eV. Reciprocal space mesh is $6\times6\times6$ for the primitive antiferromagnetic unit cell with 4 formula units. Adding a Hubbard U = 3.5 eV for the Co 3d orbital reproduces the experimental bandgap of about 2.4 eV \cite{Dudarev_1998,Agui_2011}.

The forces on atoms are relaxed to be less than $1\times10^{-3}$ eV/\AA. The relaxed lattice with antiferromagnetic order (no spin-orbit coupling) can still be described by a rhombohedral cell with the lattice constant of $5.436$ \AA\ and angle of $55.21^\circ$. This is close to the experimental lattice constant at room temperature of $5.4846$ \AA\ and angle of $55.01^\circ$ \cite{Newnham_1964}.

Phonon dispersion is obtained from force constants calculated with a $2\times2\times1$ supercell of the conventional antiferromagnetic unit cell, with phonopy\cite{phonopy-JPCM,phonopy-JPSJ}. Our calculated IR-active ($E_u$ and $A_u$) and Raman-active ($E_g$ and $A_g$) phonon frequencies are summarized below:
\begin{table}[h!]
    \centering
    \begin{tabular}{cc}
    \hline
    Phonon & Frequency (THz) \\ \hline
       $E_u$  & 6.79, 8.73, 12.34, 15.39  \\
       $A_u$  & 7.24, 11.07, 14.57, 19.98 \\
       $E_g$ & 6.28, 8.03, 9.71, 13.54, 18.18 \\
       $A_g$ & 5.11, 7.17, 11.22, 14.05, 20.34 \\ \hline
    \end{tabular}
    \caption{Calculated IR- and Raman-active phonon frequencies by DFT.}
    \label{tab:dft_phonon}
\end{table}

Magnon dispersion is calculated by SpinW implemented in MATLAB \cite{spinw}. The magnetic interaction is described by a simple XY-Heisenberg model, Yuan et. al \cite{Yuan_2020}: 
\begin{equation*}
     \mathcal{H}=\sum_{i, \delta} J_\delta^{\perp}\left(S_i^x S_{i+\delta}^x+S_i^y S_{i+\delta}^y\right)+J_\delta^z S_i^z S_{i+\delta}^z
\end{equation*}
where $J_1^\perp$ = -4.41 meV, $J_1^z$ = 0,  $J_4^\perp$ = $J_4^z$ = 0.57 meV. Here, we did not include any single-ion and exchange anisotropy, which will open a magnon gap in the acoustic branch \cite{ChoeEtAl2024PRB}, but does not qualitatively affect our experimental or theoretical results. 

Note that the zone-folding brings the acoustic phonon branches back to the zone center, which could in principle be impulsively excited and/or hybridized with the magnons at very close frequencies. However, the Raman activity of these phonons, arising from interlayer magnetoelastic coupling, should be very weak compared to that of the magnons, which have a distinctive magnetic field dependence as shown by Choe \textit{et al}~\cite{ChoeEtAl2024PRB}. Therefore, we believe that the 1.3 THz feature seen in our polarization-rotation signal based on Raman activity is primarily magnonic, although the phononic component, if any, may contribute to the excitation process from the high-frequency phonons.

\section*{Acknowledgments}
We acknowledge Martin Rodriguez-Vega for the helpful discussion regarding SpinW calculations. H.Z. acknowledges support from the National Science Foundation DMR-2240106, DMR-2327827, and the Robert A. Welch Foundation (No. C-2128). J.K. acknowledges support from the U.S. Army Research Office (through Award No. W911NF2110157), the Gordon and Betty Moore Foundation (through Grant No. 11520), and the Robert A. Welch Foundation (through Grant No. C-1509). 

\section*{References}
\bibliography{2-references}
\end{document}